\documentclass[a4paper,11pt]{article}
\usepackage{pos}

\newcommand{\cN}{{\cal N}}
\newcommand{\be}{\begin{equation}}
\newcommand{\ee}{\end{equation}}
\newcommand{\E}{\mathbb{E}}
\newcommand{\el}{\vspace*{0.3cm}}

\title{Diffusion models and stochastic quantisation in lattice field theory}
 \ShortTitle{Diffusion models and stochastic quantisation}

\author*[a]{Gert Aarts}
\author[b]{Lingxiao Wang}
\author[c,d]{Kai Zhou}

\affiliation[a]{Department of Physics, Swansea University, Swansea, SA2 8PP, United Kingdom}
\affiliation[b]{Interdisciplinary Theoretical and Mathematical Sciences Program (iTHEMS), RIKEN, Wako, Saitama 351-0198, Japan}
\affiliation[c]{School of Science and Engineering, The Chinese University of Hong Kong, Shenzhen (CUHK-Shenzhen), Guangdong, 518172, China}
\affiliation[d]{Frankfurt Institute for Advanced Studies, Ruth Moufang Strasse 1, D-60438, Frankfurt am Main, Germany\\
\mbox{}
}

\emailAdd{g.aarts@swansea.ac.uk, lingxiao.wang@riken.jp,  zhoukai@cuhk.edu.cn}

\abstract{Diffusion models are currently the leading generative AI approach used for image generation in e.g.\ DALL-E and Stable Diffusion. In this talk we relate diffusion models to stochastic quantisation in field theory and employ it to generate configurations for scalar fields on a two-dimensional lattice. We end with some speculations on possible applications. }

\FullConference{The 41st International Symposium on Lattice Field Theory (LATTICE2024)\\
 28 July - 3 August 2024\\
Liverpool, UK\\}

\begin{document}
\maketitle

\section{Introduction}

In recent years, a rich programme has been developed to apply methods of artificial intelligence and machine learning (AI/ML) to lattice field theories (LTFs), see e.g.\ Refs.~\cite{Boyda:2022nmh,Cranmer:2023xbe,Kanwar:2024ujc}. One particular direction is the use of ML to generate LFT configurations, going beyond standard approaches, such as hybrid Monte Carlo (HMC) \cite{Duane:1987de}. One reason is the notion that a well-trained ML model will generate new configurations fast, with reduced auto-correlations and possibly not suffering from critical slowing down. Evidence for this can be found in e.g.\ Refs.~\cite{Albergo:2019eim,Kanwar:2020xzo} and can intuitively be understood as follows: in a trained model each configuration is generated starting from a fresh initial configuration, which is sampled from a simple prior, rather than from a long chain of configurations with potentially lingering correlations. 
Besides this important promise, it is noted 
that LFT is an ideal playground to learn and develop ML approaches in the context of theoretical physics. 

Generally speaking, there are two schemes to devise ML algorithms to generate configurations: 
\begin{itemize}
 \item Generate configurations by approximating the (unnormalised) target distribution, $p(\phi) \sim e^{-S(\phi)}$, directly, as is done in e.g.\ normalising flow \cite{rezende2015variational,Noe:2019, Albergo:2019eim,Nicoli:2019gun,Kanwar:2020xzo,Nicoli2021,DelDebbio:2021qwf,Nicoli:2023qsl} and variations thereof, such as continuous normalizing flow~\cite{Chen:2018,deHaan:2021erb, Gerdes:2022eve,Caselle:2023mvh} and stochastic normalizing flow \cite{wu2020stochastic, Caselle:2022acb};
\item 
Approximate the underlying distribution by learning from data, i.e.\ previously generated ensembles, as is done in e.g.\  GANs \cite{Zhou:2018ill} and diffusion models, discussed here.
\end{itemize}

Recently we have introduced diffusion models in the context of LFTs. We have explored the relation between diffusion models and stochastic quantisation in scalar field theory in Refs.~\cite{Wang:2023exq,Wang:2023sry} and extended this to U(1) gauge theories in Ref.~\cite{Zhu:2024kiu}. In Ref.~\cite{Aarts:2024rsl} we studied in detail the evolution of higher-order cumulants, encoding the interactions in field theory, and compared two popular --- variance-exploding and variance-preserving or DDPM --- schemes. At this conference, we also showed first results of the application of diffusion models for theories with a complex action in which configurations are generated using complex Langevin dynamics \cite{Habibi:2024fbn}. 
Further connections between diffusion models and field theory are pointed out in Refs.~\cite{Hirono:2024zyg,Fukushima:2024oij}.
In this contribution we give a high-level overview, referring to the references above for further detail.

\section{Diffusion models and stochastic quantisation}

Diffusion models are an extremely popular approach in Generative AI, used by e.g.\ DALLE-E \cite{2022arXiv220406125R} and Stable Diffusion \cite{Rombach_2022_CVPR}. Interestingly, the method is based on concepts in non-equilibrium physics, with one of the pioneering papers called {\em Deep unsupervised learning using non-equilibrium thermodynamics} \cite{sohl-dickstein:2015deep}. Some obvious questions are:
\begin{itemize}
\item Can one use diffusion models in lattice field theory?
\item Is there a physics connection with existing methods? 
\item Is the method competitive with other approaches? 
\end{itemize}
The first two questions are answered positively in Ref.~\cite{Wang:2023exq}, which also contains encouraging indications for the third one. As mentioned, more details can be found in Refs.~\cite{Wang:2023sry,Zhu:2024kiu,Aarts:2024rsl,Habibi:2024fbn}. 

\begin{figure}[t]
    \centering
    \includegraphics[width=0.9\linewidth]{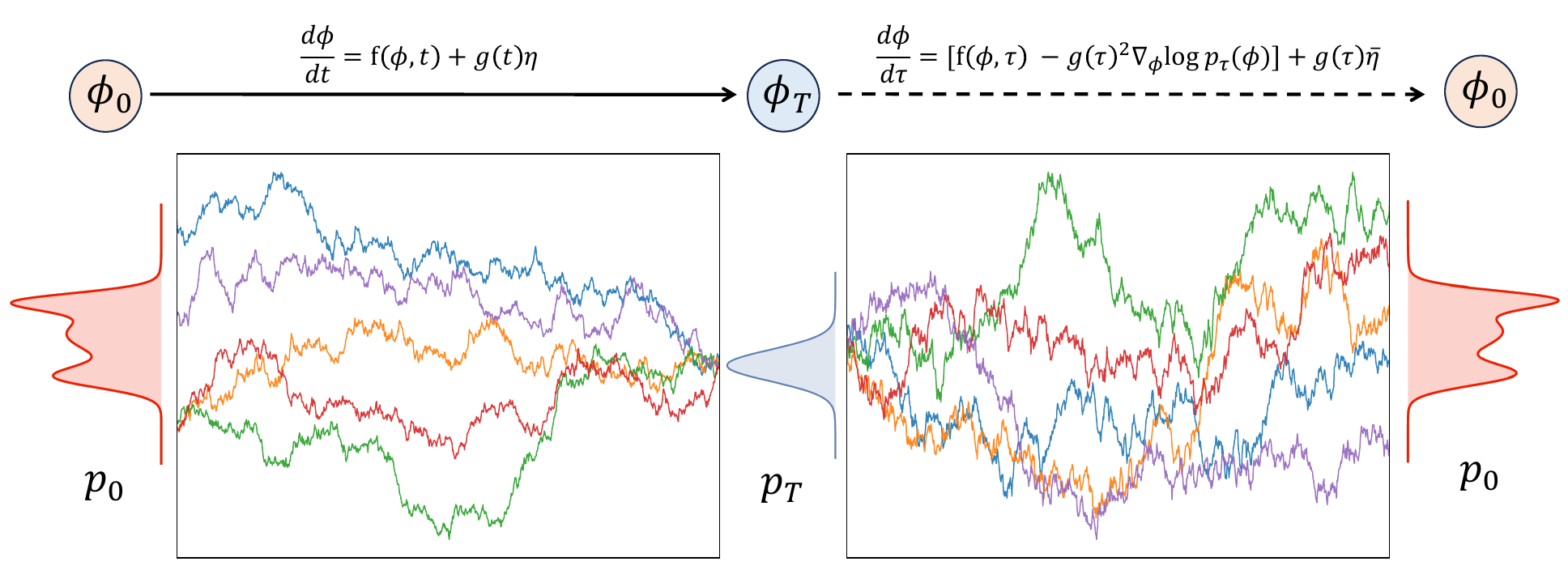}
    \caption{Sketch of the forward and backward processes in a diffusion model. The forward process, with $t= 0 \ldots T$, evolves configurations representing the target distribution $p_0$ to ones forming a simpler distribution $p_T$, while the backward process, with $\tau = T \ldots 0$,  reverses this to generate new configurations (``denoising''). The additional term in the backward process, $\nabla_\phi\log p_\tau(\phi)$, is the score, which is approximated by a neural network. From Ref.~\cite{Wang:2023exq}.
    }
    \label{fig:DM}
\end{figure}

Diffusion models work in combination with a previously obtained set of images or configurations, representing the target distribution $p_0(\phi)$. During the forward process, these images are made blurry or noisy, using a stochastic process. During the backward process, this is reversed and new images or configurations are created (``denoising''), starting from a normal distribution. This setup is illustrated in Fig.~\ref{fig:DM}.
The crucial difference between the forward and backward process is the presence of the so-called score, $\nabla_\phi\log p_\tau(\phi)$, which controls the convergence of the backward process. The score is approximated by a neural network and learnt during the forward process.

In the simplest case, with no drift applied during the forward process (i.e.\ $f(\phi,t)=0$ in Fig.~\ref{fig:DM}), the stochastic equations read
\begin{align}
\label{eq1}
&\mbox{forward:} \quad &&\partial_t \phi(x,t)  = g(t)\eta(x,t), \\
\label{eq2}
&\mbox{backward:}  \quad &&\partial_\tau\phi(x,\tau) = g^2(T-\tau)\nabla_\phi \log p(\phi,T-\tau) + g(T-\tau)\eta(x,\tau).
\end{align}
Here $\eta(x,t)\sim \cN(0,1)$ is Gaussian noise with variance 1, applied locally at each pixel or lattice coordinate, and $g(t)$ is the diffusion coefficient, setting the time-dependent noise strength. A common choice is $g(t)=\sigma^{t/T}$, with $\sigma\gg 1$. Compared to Fig.~\ref{fig:DM}, we have redefined time in the backward process, $\tau\to T-\tau$, such that $0\leq t, \tau \leq T$. Importantly, the time intervals are finite. 
The scheme with no drift, as in Eqs.~(\ref{eq1}, \ref{eq2}), is commonly referred to as the variance-exploding scheme, since the variance increases in time as $\E[\phi^2(x,t)] \sim \sigma^{2t/T}$, such that the noise will dominate the signal at the end of the forward process.
A detailed analysis of the evolution of the higher-order moments and cumulants can be found in Ref.~\cite{Aarts:2024rsl}.

\begin{figure}[t]
    \centering
    \includegraphics[width=0.8\linewidth]{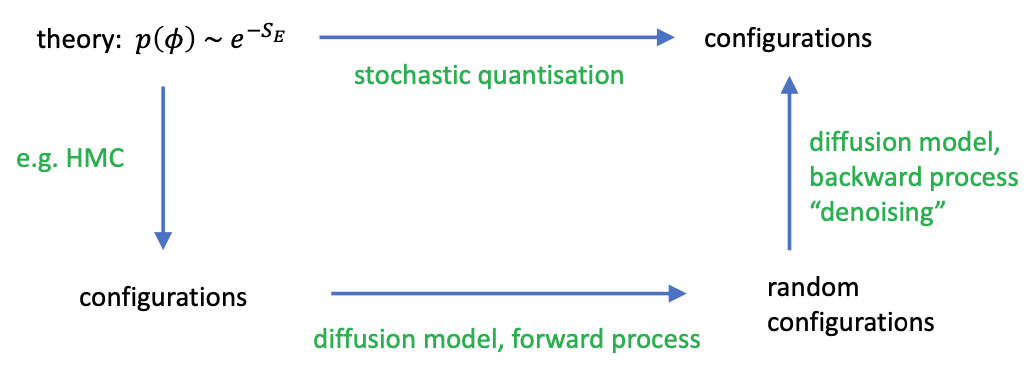}
    \caption{
    Flow chart indicating the relation between stochastic quantisation and diffusion models in the case of lattice field theory. Starting from the defining theory (top left), new configurations (top right) can be generated via stochastic quantisation or via the application of a diffusion model, trained on pre-existing configurations (bottom left). 
    From Ref.~\cite{Wang:2023exq}.
  }
    \label{fig:flowchart}
\end{figure}

If we assume that the score follows from a time-dependent distribution, 
\be
 p(\phi,t) = \frac{1}{Z} \exp[-S(\phi,t)] \qquad \Rightarrow \qquad
\nabla_\phi \log p(\phi,t) = -\nabla S(\phi,t),
\ee
the backward process takes a familiar form
\be
\partial_\tau\phi(x,\tau) = -g^2(T-\tau)\nabla S(\phi,T-\tau) + g(T-\tau)\eta(x,\tau).
\ee
One cannot help but notice that this is similar to the equation encountered in stochastic quantisation, i.e., path integral quantisation via a stochastic process in a fictitious time \cite{Parisi:1980ys,Damgaard:1987rr},
\be
\partial_\tau\phi(x,\tau) = -\nabla S(\phi,\tau) + \sqrt{2}\eta(x,\tau).
\ee
Besides the normalisation of the noise (which can be changed by rescaling the time step), we note the following: 
\begin{itemize}
\item {\bf stochastic quantisation}: \\
-- the drift is time-independent and determined by a known action; \\
-- the noise variance is constant (but this can be generalised using kernels \cite{Damgaard:1987rr}\footnote{Kernels change the dynamics but leave the stationary solution unchanged; for a recent application, 
see e.g.\ Ref.~\cite{Alvestad:2022abf}.}); \\
-- the dynamics consists of a thermalisation stage followed by evolution in equilibrium during which measurements are made.
\item {\bf diffusion models}: \\
-- the drift or score is not known a priori but is learnt from data; \\
-- the score and diffusion coefficient are time-dependent; \\
-- the evolution consists of many short runs ($0\leq \tau\leq T$), 
with measurements taken at $\tau=T$; \\
-- correlations between runs starting from a simple prior should be absent and generated configurations can be used as proposals in a Markov chain with reduced auto-correlation.
\end{itemize}
A flow chart summarising the relation between stochastic quantisation and diffusion models is given in Fig.~\ref{fig:flowchart}. If all algorithms are working well, the ensembles generated bottom left and top right are all representative of the target distribution $p(\phi)\sim \exp[-S(\phi)]$.

\section{Two-dimensional scalar fields}

\begin{figure}[t]
    \centering
    \includegraphics[width=0.7\linewidth]{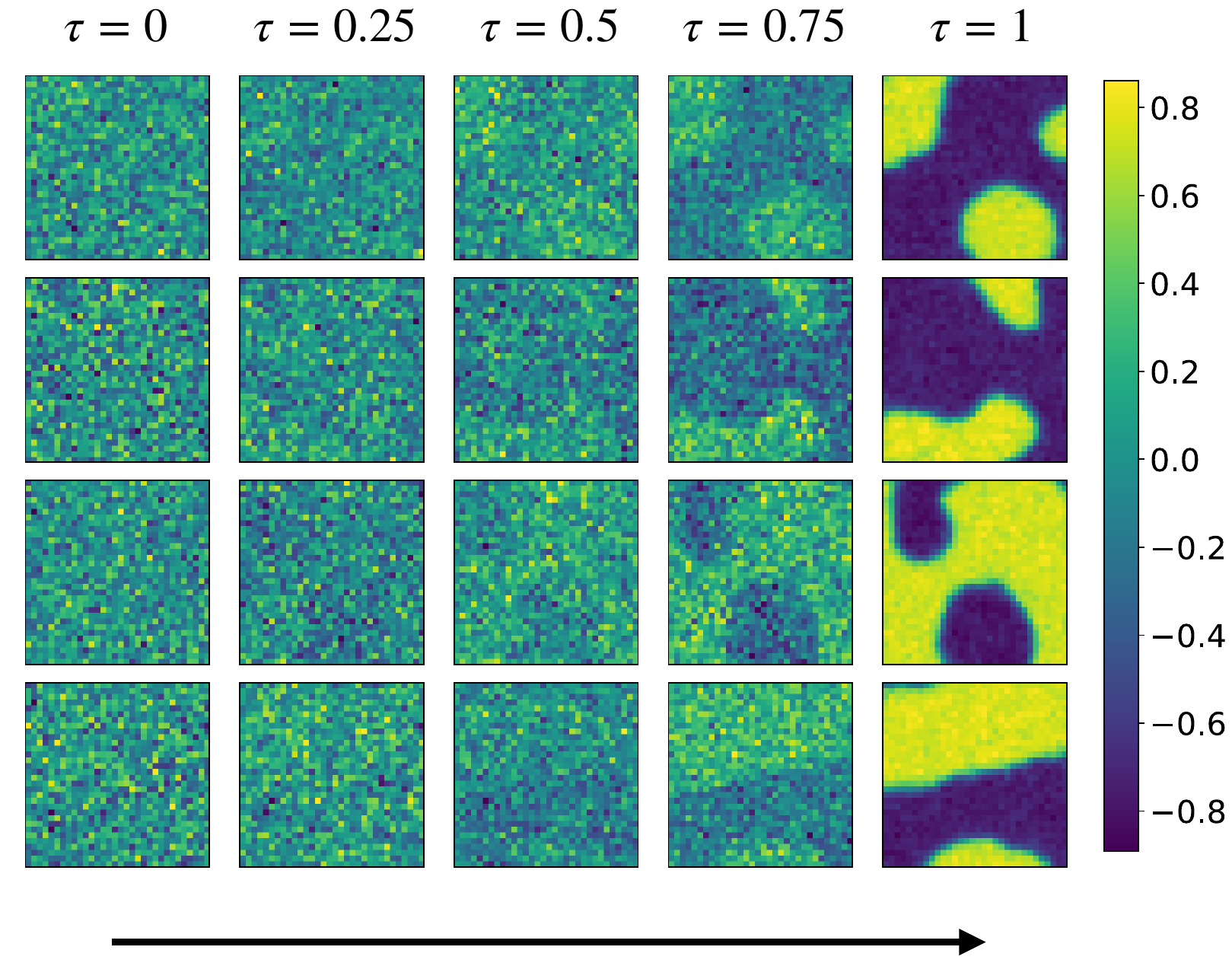}
    \caption{Denoising process in a two-dimensional scalar field theory. Four independent configurations are generated at the end of the backward process ($\tau=1$), with clusters characteristic of the broken phase. From Ref.~\cite{Wang:2023exq}.
    }
    \label{fig:clusters}
\end{figure}

We have applied the diffusion model in the variance-exploding scheme to a $\lambda\phi^4$ theory on a two-dimensional lattice, with parameter choices in the symmetric and broken phase \cite{Wang:2023exq}. The results shown here are obtained on a volume of $32^2$. Training data is generated using Hybrid Monte Carlo; for the results shown here the diffusion model was trained using a U-net architecture on ensembles with 5120 independent configurations. Fig.~\ref{fig:clusters} shows the denoising process in action during the backward process: four independent configurations are generated, with clusters characteristic of the broken phase appearing at the end of the backward process ($\tau=1$). For a detailed discussion, including comparisons of the susceptibility, the Binder cumulant and higher-order cumulants, as well as correlation times and acceptance rates, we refer to Refs.~\cite{Wang:2023exq,Aarts:2024rsl}.

\begin{figure}[t]
    \centering
    \includegraphics[width=0.8\linewidth]{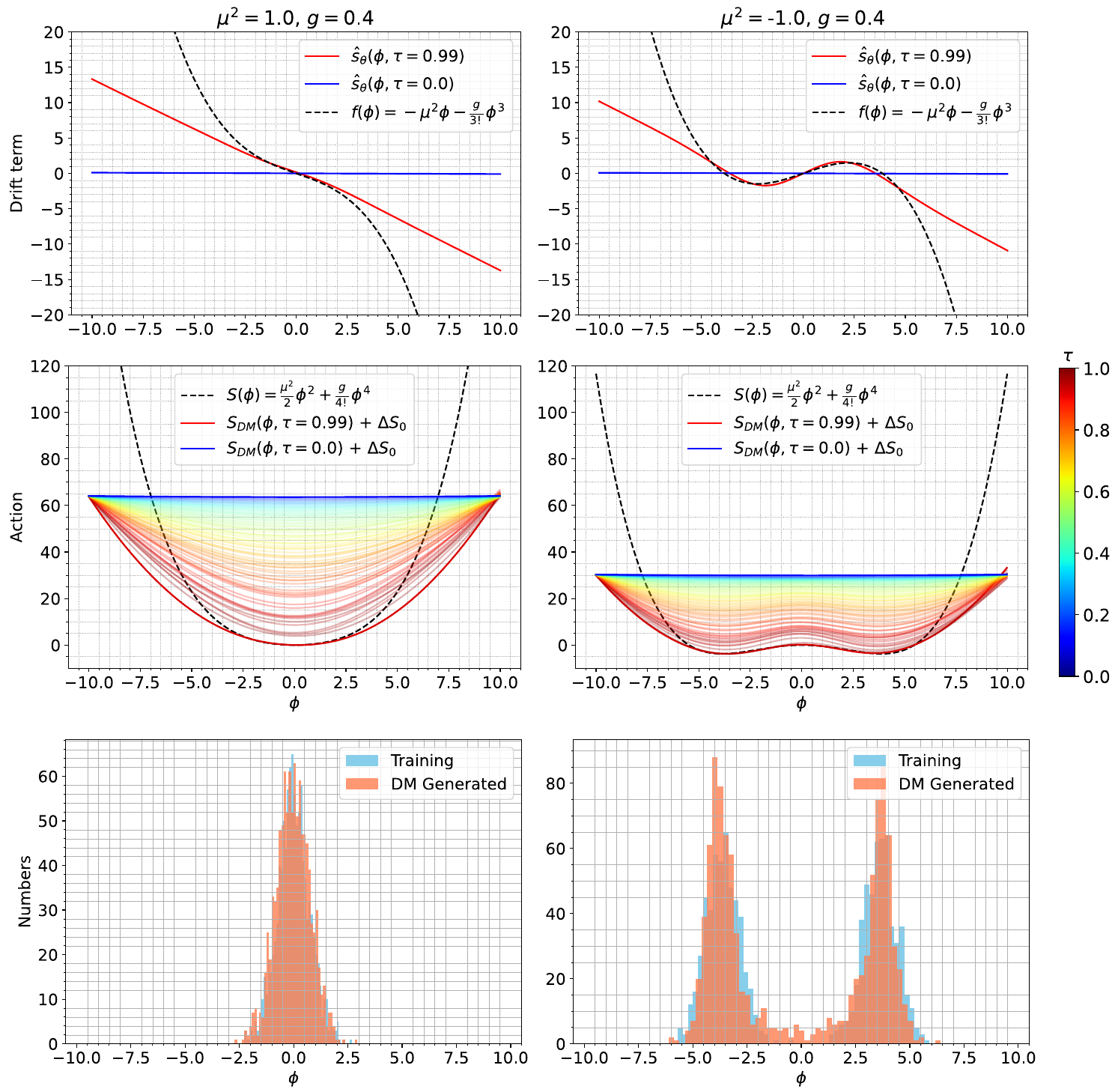}
    \caption{
    Toy model: 
    Drift terms (upper row) and effective actions (middle row) learned by the diffusion model as a function of $\phi$ in both single-well (left column) and double-well (right column) actions, for various values of the time $\tau$ during the stochastic process. The action is shifted by a constant $\Delta S_0$. The dashed lines indicate the exact values. The bottom row shows 1024 samples generated using the target distribution and the trained diffusion model.
   From Ref.~\cite{Wang:2023exq}. }
    \label{fig:toymodel}
\end{figure}

To illustrate how the diffusion model interpolates between the prior and the target distribution, we show in Fig.~\ref{fig:toymodel} the evolution during the backward process of the drift (top row) and the time-dependent action (middle row) learnt by the diffusion model in the case of a simple model with one degree of freedom, with the action
\be
S(\phi) =  \frac{1}{2}\mu^2 \phi^2 + \frac{1}{4!}g \phi^4, \quad\qquad \mu^2 = \pm 1, g=0.4.
\ee
The dashed lines indicate the exact (target) results, and coloured lines show the evolution from $\tau=0$ (blue) to $\tau=1$ (ref). The bottom row finally shows samples generated directly from the target distribution and from the trained diffusion model. It is worth pointing out that the diffusion model can only learn where data is available, which explains the deviations seen for larger values of $|\phi|$ in the top and middle rows.

\section{Outlook}

In this contribution we have only shown the start of a programme to apply diffusion models to generate configurations in lattice field theory and supplement existing ensembles. Indeed, directions to go into are plenty. Gauge theories can be included combining insights from stochastic quantisation and gauge-equivariant networks \cite{Kanwar:2020xzo,Boyda:2020hsi,Favoni:2020reg}. The first application to a U(1) gauge theory can be found in Ref.~\cite{Zhu:2024kiu}.
Fermions can be included implicitly, with their presence imprinted on bosonic field configurations generated in theories with fermions. An interesting direction is to apply diffusion models to theories with a sign or complex action problem, learning the (real and semi-positive) distribution from configurations generated by complex Langevin dynamics, which is not known a priori \cite{Parisi:1983mgm,Aarts:2008rr,Seiler:2012wz,Aarts:2015tyj}. This is further discussed in Ref.~\cite{Habibi:2024fbn}. Finally, in all cases it is important to make the algorithm exact, by including an efficient accept-reject step, and demonstrating an improvement over existing algorithms, e.g.\ by evading critical slowing down.
Work in all these directions is currently in progress.

\el 

\noindent
{\bf Acknowledgements} --  
We thank ECT* and the ExtreMe Matter Institute EMMI at GSI, Darmstadt, for support during the ECT*/EMMI workshop {\em Machine learning for lattice field theory and beyond} in June 2023. 
GA is supported by STFC Consolidated Grant ST/T000813/1. 
LW thanks the DEEP-IN working group at RIKEN-iTHEMS for support.
KZ is supported by the CUHK-Shenzhen University development fund under grant No.\ UDF01003041 and UDF03003041, and Shenzhen Peacock fund under No.\ 2023TC0179.

\noindent
{\bf Research Data and Code Access} --
Details of the code and data presented  can be found in Ref.~\cite{Wang:2023exq}.

\noindent
{\bf Open Access Statement} -- For the purpose of open access, the authors have applied a Creative Commons Attribution (CC BY) licence to any Author Accepted Manuscript version arising.

\providecommand{\href}[2]{#2}\begingroup\raggedright\endgroup


\begin{thebibliography}{10}

\bibitem{Boyda:2022nmh}
D.~Boyda et~al., \emph{{Applications of Machine Learning to Lattice Quantum
  Field Theory}},  in \emph{{Snowmass 2021}}, 2022
  [\href{https://arxiv.org/abs/2202.05838}{{\ttfamily 2202.05838}}].

\bibitem{Cranmer:2023xbe}
K.~Cranmer, G.~Kanwar, S.~Racani\`ere, D.J.~Rezende and P.E.~Shanahan,
  \emph{{Advances in machine-learning-based sampling motivated by lattice
  quantum chromodynamics}},
  \href{https://doi.org/10.1038/s42254-023-00616-w}{\emph{Nature Rev. Phys.}
  {\bfseries 5} (2023) 526} [\href{https://arxiv.org/abs/2309.01156}{{\ttfamily
  2309.01156}}].

\bibitem{Kanwar:2024ujc}
G.~Kanwar, \emph{{Flow-based sampling for lattice field theories}},  in
  \emph{{40th International Symposium on Lattice Field Theory}}, 2024
  [\href{https://arxiv.org/abs/2401.01297}{{\ttfamily 2401.01297}}].

\bibitem{Duane:1987de}
S.~Duane, A.D.~Kennedy, B.J.~Pendleton and D.~Roweth, \emph{{Hybrid Monte
  Carlo}}, \href{https://doi.org/10.1016/0370-2693(87)91197-X}{\emph{Phys.
  Lett. B} {\bfseries 195} (1987) 216}.

\bibitem{Albergo:2019eim}
M.S.~Albergo, G.~Kanwar and P.E.~Shanahan, \emph{Flow-based generative models
  for {{Markov}} chain {{Monte Carlo}} in lattice field theory},
  \href{https://doi.org/10.1103/PhysRevD.100.034515}{\emph{Phys. Rev. D}
  {\bfseries 100} (2019) 034515}
  [\href{https://arxiv.org/abs/1904.12072}{{\ttfamily 1904.12072}}].

\bibitem{Kanwar:2020xzo}
G.~Kanwar, M.S.~Albergo, D.~Boyda, K.~Cranmer, D.C.~Hackett, S.~Racani{\`e}re
  et~al., \emph{Equivariant {{Flow-Based Sampling}} for {{Lattice Gauge
  Theory}}}, \href{https://doi.org/10.1103/PhysRevLett.125.121601}{\emph{Phys.
  Rev. Lett.} {\bfseries 125} (2020) 121601}
  [\href{https://arxiv.org/abs/2003.06413}{{\ttfamily 2003.06413}}].

\bibitem{rezende2015variational}
D.J.~Rezende and S.~Mohamed, \emph{{Variational Inference with Normalizing
  Flows}},  in \emph{International conference on machine learning},
  pp.~1530--1538, PMLR, 2015
  [\href{https://arxiv.org/abs/1505.05770}{{\ttfamily 1505.05770}}].

\bibitem{Noe:2019}
F.~Noé, S.~Olsson, J.~Köhler and H.~Wu, \emph{Boltzmann generators: Sampling
  equilibrium states of many-body systems with deep learning},
  \href{https://doi.org/10.1126/science.aaw1147}{\emph{Science} {\bfseries 365}
  (2019) eaaw1147} [\href{https://arxiv.org/abs/1812.01729}{{\ttfamily
  1812.01729}}].

\bibitem{Nicoli:2019gun}
K.A.~Nicoli, S.~Nakajima, N.~Strodthoff, W.~Samek, K.-R.~M\"uller and
  P.~Kessel, \emph{{Asymptotically unbiased estimation of physical observables
  with neural samplers}},
  \href{https://doi.org/10.1103/PhysRevE.101.023304}{\emph{Phys. Rev. E}
  {\bfseries 101} (2020) 023304}
  [\href{https://arxiv.org/abs/1910.13496}{{\ttfamily 1910.13496}}].

\bibitem{Nicoli2021}
K.A.~Nicoli, C.J.~Anders, L.~Funcke, T.~Hartung, K.~Jansen, P.~Kessel et~al.,
  \emph{Estimation of thermodynamic observables in lattice field theories with
  deep generative models},
  \href{https://doi.org/10.1103/PhysRevLett.126.032001}{\emph{Phys. Rev. Lett.}
  {\bfseries 126} (2021) 032001}
  [\href{https://arxiv.org/abs/2007.07115}{{\ttfamily 2007.07115}}].

\bibitem{DelDebbio:2021qwf}
L.~Del~Debbio, J.~Marsh~Rossney and M.~Wilson, \emph{Efficient modeling of
  trivializing maps for lattice $\phi^4$ theory using normalizing flows: {{A}}
  first look at scalability},
  \href{https://doi.org/10.1103/PhysRevD.104.094507}{\emph{Phys. Rev. D}
  {\bfseries 104} (2021) 094507}
  [\href{https://arxiv.org/abs/2105.12481}{{\ttfamily 2105.12481}}].

\bibitem{Nicoli:2023qsl}
K.A.~Nicoli, C.J.~Anders, T.~Hartung, K.~Jansen, P.~Kessel and S.~Nakajima,
  \emph{{Detecting and Mitigating Mode-Collapse for Flow-based Sampling of
  Lattice Field Theories}},  \href{https://arxiv.org/abs/2302.14082}{{\ttfamily
  2302.14082}}.

\bibitem{Chen:2018}
R.T.~Chen, Y.~Rubanova, J.~Bettencourt and D.K.~Duvenaud, \emph{Neural ordinary
  differential equations}, {\emph{Advances in neural information processing
  systems} {\bfseries 31} (2018) }
  [\href{https://arxiv.org/abs/1806.07366}{{\ttfamily 1806.07366}}].

\bibitem{deHaan:2021erb}
P.~de~Haan, C.~Rainone, M.C.N.~Cheng and R.~Bondesan, \emph{{Scaling Up Machine
  Learning For Quantum Field Theory with Equivariant Continuous Flows}},
  \href{https://arxiv.org/abs/2110.02673}{{\ttfamily 2110.02673}}.

\bibitem{Gerdes:2022eve}
M.~Gerdes, P.~de~Haan, C.~Rainone, R.~Bondesan and M.C.N.~Cheng,
  \emph{{Learning lattice quantum field theories with equivariant continuous
  flows}}, \href{https://doi.org/10.21468/SciPostPhys.15.6.238}{\emph{SciPost
  Phys.} {\bfseries 15} (2023) 238}
  [\href{https://arxiv.org/abs/2207.00283}{{\ttfamily 2207.00283}}].

\bibitem{Caselle:2023mvh}
M.~Caselle, E.~Cellini and A.~Nada, \emph{{Sampling the lattice Nambu-Goto
  string using Continuous Normalizing Flows}}, {\emph{Journal of High Energy
  Physics} {\bfseries 02} (2024) 048}
  [\href{https://arxiv.org/abs/2307.01107}{{\ttfamily 2307.01107}}].

\bibitem{wu2020stochastic}
H.~Wu, J.~K{\"o}hler and F.~No{\'e}, \emph{Stochastic normalizing flows},
  {\emph{Advances in Neural Information Processing Systems} {\bfseries 33}
  (2020) 5933} [\href{https://arxiv.org/abs/2002.06707}{{\ttfamily
  2002.06707}}].

\bibitem{Caselle:2022acb}
M.~Caselle, E.~Cellini, A.~Nada and M.~Panero, \emph{Stochastic normalizing
  flows as non-equilibrium transformations},
  \href{https://doi.org/10.1007/JHEP07(2022)015}{\emph{JHEP} {\bfseries 07}
  (2022) 015} [\href{https://arxiv.org/abs/2201.08862}{{\ttfamily
  2201.08862}}].

\bibitem{Zhou:2018ill}
K.~Zhou, G.~Endr{\H o}di, L.-G.~Pang and H.~St{\"o}cker, \emph{Regressive and
  generative neural networks for scalar field theory},
  \href{https://doi.org/10.1103/PhysRevD.100.011501}{\emph{Phys. Rev. D}
  {\bfseries 100} (2019) 011501}.

\bibitem{Wang:2023exq}
L.~Wang, G.~Aarts and K.~Zhou, \emph{{Diffusion models as stochastic
  quantization in lattice field theory}},
  \href{https://doi.org/10.1007/JHEP05(2024)060}{\emph{JHEP} {\bfseries 05}
  (2024) 060} [\href{https://arxiv.org/abs/2309.17082}{{\ttfamily
  2309.17082}}].

\bibitem{Wang:2023sry}
L.~Wang, G.~Aarts and K.~Zhou, \emph{{Generative Diffusion Models for Lattice
  Field Theory}},  in \emph{{37th Conference on Neural Information Processing
  Systems}}, 2023 [\href{https://arxiv.org/abs/2311.03578}{{\ttfamily
  2311.03578}}].

\bibitem{Zhu:2024kiu}
Q.~Zhu, G.~Aarts, W.~Wang, K.~Zhou and L.~Wang, \emph{{Diffusion models for
  lattice gauge field simulations}},  in \emph{{38th conference on Neural
  Information Processing Systems}}, 2024
  [\href{https://arxiv.org/abs/2410.19602}{{\ttfamily 2410.19602}}].

\bibitem{Aarts:2024rsl}
G.~Aarts, D.E.~Habibi, L.~Wang and K.~Zhou, \emph{{On learning higher-order
  cumulants in diffusion models}},  in \emph{{38th conference on Neural
  Information Processing Systems}}, 2024
  [\href{https://arxiv.org/abs/2410.21212}{{\ttfamily 2410.21212}}].

\bibitem{Habibi:2024fbn}
D.E.~Habibi, G.~Aarts, L.~Wang and K.~Zhou, \emph{{Diffusion models learn
  distributions generated by complex Langevin dynamics}},  in \emph{{41st
  International Symposium on Lattice Field Theory}}, 2024
  [\href{https://arxiv.org/abs/2412.01919}{{\ttfamily 2412.01919}}].

\bibitem{Hirono:2024zyg}
Y.~Hirono, A.~Tanaka and K.~Fukushima, \emph{{Understanding Diffusion Models by
  Feynman's Path Integral}},
  \href{https://arxiv.org/abs/2403.11262}{{\ttfamily 2403.11262}}.

\bibitem{Fukushima:2024oij}
K.~Fukushima and S.~Kamata, \emph{{Stochastic quantization and diffusion
  models}},  \href{https://arxiv.org/abs/2411.11297}{{\ttfamily 2411.11297}}.

\bibitem{2022arXiv220406125R}
A.~{Ramesh}, P.~{Dhariwal}, A.~{Nichol}, C.~{Chu} and M.~{Chen},
  \emph{{Hierarchical Text-Conditional Image Generation with CLIP Latents}},
  \href{https://arxiv.org/abs/2204.06125}{{\ttfamily 2204.06125}}.

\bibitem{Rombach_2022_CVPR}
R.~Rombach, A.~Blattmann, D.~Lorenz, P.~Esser and B.~Ommer,
  \emph{{High-Resolution Image Synthesis with Latent Diffusion Models}},
  {\emph{{Proceedings of the IEEE/CVF Conference on Computer Vision and Pattern
  Recognition (CVPR)}} (2022) 10684}
  [\href{https://arxiv.org/abs/2112.10752}{{\ttfamily 2112.10752}}].

\bibitem{sohl-dickstein:2015deep}
J.~{Sohl-Dickstein}, E.A.~Weiss, N.~Maheswaranathan and S.~Ganguli, \emph{Deep
  unsupervised learning using nonequilibrium thermodynamics},  in \emph{Proc.
  32nd {{Int}}. {{Conf}}. {{Int}}. {{Conf}}. {{Mach}}. {{Learn}}. - {{Vol}}.
  37}, pp.~2256--2265, 2015 [\href{https://arxiv.org/abs/1503.03585}{{\ttfamily
  1503.03585}}].

\bibitem{Parisi:1980ys}
G.~Parisi and Y.S.~Wu, \emph{{Perturbation theory without gauge fixing}},
  {\emph{Sci. China, A} {\bfseries 24} (1980) 483}.

\bibitem{Damgaard:1987rr}
P.H.~Damgaard and H.~H{\"u}ffel, \emph{Stochastic quantization},
  \href{https://doi.org/10.1016/0370-1573(87)90144-X}{\emph{Phys. Rept.}
  {\bfseries 152} (1987) 227}.

\bibitem{Alvestad:2022abf}
D.~Alvestad, R.~Larsen and A.~Rothkopf, \emph{{Towards learning optimized
  kernels for complex Langevin}},
  \href{https://doi.org/10.1007/JHEP04(2023)057}{\emph{JHEP} {\bfseries 04}
  (2023) 057} [\href{https://arxiv.org/abs/2211.15625}{{\ttfamily
  2211.15625}}].

\bibitem{Boyda:2020hsi}
D.~Boyda, G.~Kanwar, S.~Racani\`ere, D.J.~Rezende, M.S.~Albergo, K.~Cranmer
  et~al., \emph{{Sampling using $SU(N)$ gauge equivariant flows}},
  \href{https://doi.org/10.1103/PhysRevD.103.074504}{\emph{Phys. Rev. D}
  {\bfseries 103} (2021) 074504}
  [\href{https://arxiv.org/abs/2008.05456}{{\ttfamily 2008.05456}}].

\bibitem{Favoni:2020reg}
M.~Favoni, A.~Ipp, D.I.~M{\"u}ller and D.~Schuh, \emph{Lattice {{Gauge
  Equivariant Convolutional Neural Networks}}},
  \href{https://doi.org/10.1103/PhysRevLett.128.032003}{\emph{Phys. Rev. Lett.}
  {\bfseries 128} (2022) 032003}
  [\href{https://arxiv.org/abs/2012.12901}{{\ttfamily 2012.12901}}].

\bibitem{Parisi:1983mgm}
G.~Parisi, \emph{On complex probabilities},
  \href{https://doi.org/10.1016/0370-2693(83)90525-7}{\emph{Physics Letters B}
  {\bfseries 131} (1983) 393}.

\bibitem{Aarts:2008rr}
G.~Aarts and I.-O.~Stamatescu, \emph{{Stochastic quantization at finite
  chemical potential}},
  \href{https://doi.org/10.1088/1126-6708/2008/09/018}{\emph{JHEP} {\bfseries
  09} (2008) 018} [\href{https://arxiv.org/abs/0807.1597}{{\ttfamily
  0807.1597}}].

\bibitem{Seiler:2012wz}
E.~Seiler, D.~Sexty and I.-O.~Stamatescu, \emph{{Gauge cooling in complex
  Langevin for QCD with heavy quarks}},
  \href{https://doi.org/10.1016/j.physletb.2013.04.062}{\emph{Phys. Lett. B}
  {\bfseries 723} (2013) 213}
  [\href{https://arxiv.org/abs/1211.3709}{{\ttfamily 1211.3709}}].

\bibitem{Aarts:2015tyj}
G.~Aarts, \emph{{Introductory lectures on lattice QCD at nonzero baryon
  number}}, \href{https://doi.org/10.1088/1742-6596/706/2/022004}{\emph{J.
  Phys. Conf. Ser.} {\bfseries 706} (2016) 022004}
  [\href{https://arxiv.org/abs/1512.05145}{{\ttfamily 1512.05145}}].

\end{thebibliography}
\end{document}